\documentclass[10pt,conference]{IEEEtran}\IEEEoverridecommandlockouts
\usepackage{epsfig,graphicx,subfigure,psfrag,amsmath,cases}
\usepackage{latexsym,amssymb,amsmath,epsfig,subfigure,algorithm}
\usepackage{subfigure}
\usepackage{bbding}
\usepackage{multicol}
\usepackage{graphics}
\usepackage{amsthm}
\usepackage{algorithmic}
\usepackage{color}
\usepackage{url}
\usepackage{scrtime}
\usepackage{bm}
\usepackage{graphicx}

\author{Yuanxin Cai$^*$, Zhiqiang Wei$^*$, Ruide Li$^\dag$, Derrick Wing Kwan Ng\thanks{ D. W. K. Ng is supported by the Australian Research Council's Discovery Early Career Researcher Award (DE170100137) and Discovery Project (DP190101363). }$^*$, and Jinhong Yuan\thanks{J. Yuan is supported by the Australia Research Council Discovery Project under Grant (DP160104566) and Linkage Project under Grant (LP 160100708).}$^*$\\
$^*$School of Electrical Engineering and Telecommunications, The University of New South Wales, Sydney, Australia\\
$^\dag$School of Information and Electronics, Beijing Institute of Technology, Beijing, China
 }

\title{Energy-Efficient Resource Allocation for Secure UAV Communication Systems}

\newtheorem{T-Prob}{Transformed Problem}

\newtheorem{Remark}{Remark}

\begin{document}
\maketitle
\begin{abstract}
  In this paper, we study the resource allocation and trajectory design for energy-efficient secure unmanned aerial vehicle (UAV) communication systems where a UAV base station serves multiple legitimate ground users in the existence of a potential eavesdropper.
  We aim to maximize the energy efficiency of the UAV by jointly optimizing its transmit power, user scheduling, trajectory, and velocity.
  The design is formulated as a non-convex optimization problem taking into account   the maximum tolerable signal-to-noise ratio (SNR) leakage, the minimum data rate requirement of each user,  and the location uncertainty of the eavesdropper.
  An iterative algorithm is proposed to obtain an efficient suboptimal solution.
  Simulation results demonstrate that the proposed algorithm can achieve a significant improvement of the system energy efficiency while satisfying communication security constraint, compared to some simple scheme adopting straight flight trajectory with a constant speed.
\end{abstract}

\renewcommand{\baselinestretch}{0.93}
\large\normalsize
\section{Introduction}

Due to the high flexibility and mobility of unmanned aerial vehicles (UAVs) offered to wireless communication systems, several interesting applications of UAV have been proposed \cite{opportunities_and_challenges}, such as mobile base stations \cite{DWK_2018_solar_power_comm}, mobile relays \cite{relays}, and mobile data collections \cite{sensor_network}, etc.
In practice, the total energy budget for maintaining stable flight and communication is limited by the onboard battery capacity.
Hence,  energy efficiency  has become an important figure of merit for UAV-based communications.
For example, the authors in \cite{sensor_network} studied the energy efficiency maximization for wireless sensor networks via jointly optimizing the weak up schedule of sensor nodes and UAV's trajectory.
Yet, the flight power consumption of the system was not considered which contributes a significant portion of total power consumption in the systems.
Besides, a UAV trajectory design was developed to optimize the system energy efficiency in \cite{EE_fixed_wing}.
However, the investigation of variable speed as well as transmit power allocation strategy for communications were not conducted which plays an important role for the design of energy-efficient UAV systems.
In \cite{jawhar2013data}, the authors compared the delivery ratio and average delay of UAV-based wireless communication systems with constant speed, variable speed, and adaptive speed of the UAV, for reducing the system energy consumption.
Yet, the study was limited to the case of multiple sensors deployed in a specific environment and their results cannot be applied to the general case with different system topologies.
Furthermore, although orthogonal frequency division multiple access (OFDMA) has been commonly adopted in conventional communication systems, an energy-efficient trajectory and resource allocation design enabling secure UAV-OFDMA wireless communication systems has not been reported in the literature yet.
Meanwhile, since the line-of-sight (LoS) paths dominate the air-to-ground communication channels, UAV-based communications are susceptible to potential eavesdropping.
Thus, there is an emerging need for designing secure UAV-based communication.
For instance, the authors in \cite{Zhang2018Securing} proposed a joint power allocation and trajectory design to maximize the secrecy rate in both uplink and downlink systems.
In  \cite{secrecy_EE},  secure energy efficiency maximization for UAV-based relaying systems was studied. However, both works only considered the case of single-user and the proposed designs in \cite{Zhang2018Securing,secrecy_EE} are not applicable to the case of multiple users.
Besides, the availability of the eavesdropper location was assumed in \cite{Zhang2018Securing,secrecy_EE}, which is generally over optimistic.
Although \cite{cui2018robust} studied the resource allocation design for secure UAV systems by taking into account the imperfect channel state information (CSI) of an eavesdropper, the energy efficiency of such systems is still an unknown.
In this paper, we tackle the aforementioned problems via optimizing the trajectory and resource allocation strategy for energy-efficient secure UAV-OFDMA systems with multiple legitimate users and the existence of a potential eavesdropper.
Particularly, the malicious eavesdropper is located at an uncertain region between the UAV's initial location and its destination.
By exploiting the high flexibility of the UAV, one can either reduce its transmit power or fly away from the uncertain region centered at the eavesdropper to guarantee secure communications for legitimate users.
We aim to propose a joint design of resource allocation and trajectory to maximize the system energy efficiency while considering  the maximum tolerable signal leakage of the eavesdropper and the minimum individual user data rate requirement.
An iterative algorithm is proposed to achieve a suboptimal solution of the design problem.
Simulation results unveil that the performance of our proposed algorithm offers a considerable system  energy efficiency  compared to a baseline scheme adopting a straight flight trajectory and a constant speed.
Notation:
$\mathbb{R}^{M \times 1}$ is the space of a M-dimensional real-valued vector.
$\|\cdot\|$ denotes the vector norm.
$\mathbf{I}_n$ represents an $n \times n$ identity matrix.
$[x]^+ = \max \{0,x\}$. $[\cdot]^{\mathrm{T}}$ denotes the transpose operation.
For a vector $\mathbf{a}$, $\|\mathbf{a}\|$ represents its  norm.

\section{System Model}

\begin{figure}[t]\vspace*{2mm}
  \centering
  \includegraphics[width=3.5 in]{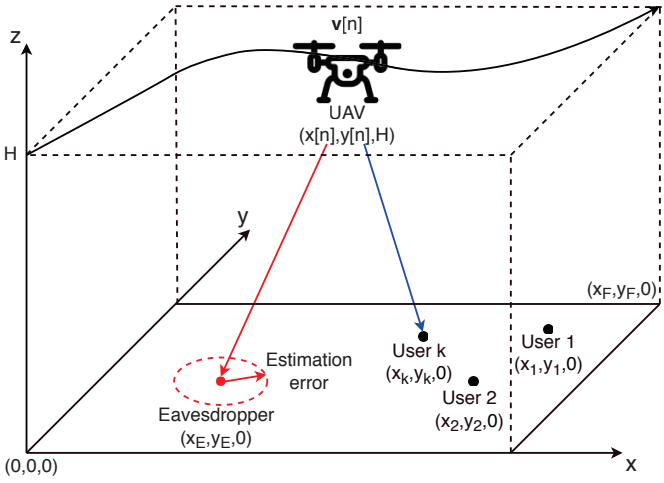}
  \caption{An UAV-OFDMA system with multiple users and a potential eavesdropper with  location uncertainty.}
  \label{system_model}
\end{figure}

A UAV-based OFDMA communication system is considered which consists of a UAV serving as a transmitter, $K$ legitimate users, and a potential eavesdropper, as shown in Fig. \ref{system_model}.
All the transmitter and receivers are single-antenna devices.
We assume that the total bandwidth and the time duration of the system are divided equally into $N_{\mathrm{F}}$ subcarriers and $N$ time slots, respectively.
In this system, we assume that the UAV flies at a constant altitude $H$ and  all the ground nodes remain steady for $N$ time slots.
The distance between the UAV and user $k\in\{1,\ldots,K\}$ at time slot $n\in\{1,\ldots,N\}$ is given by
\begin{eqnarray}
d_k[n] = \sqrt{\|\mathbf{t}_k-\mathbf{t}[n]\|^2 +H^2},
\end{eqnarray}
where $\mathbf{t}_k = [x_k,y_k]^{\mathrm{T}} \in \mathbb{R}^{2 \times 1} $ represents the location of the ground user $k$, and $\mathbf{t}[n] = [x[n], y[n]]^{\mathrm{T}} \in \mathbb{R}^{2 \times 1}$ represents the horizontal location of the UAV at time slot $n$.
Similarly, the distance between the UAV and the potential eavesdropper at time slot $n$ can be modeled by
\begin{eqnarray}
d_{\mathrm{E}}[n] = \sqrt{\| \hat{\mathbf{t}}_{\mathrm{E}} + \Delta \mathbf{t}_{\mathrm{E}} - \mathbf{t}[n] \|^2 + H^2},
\end{eqnarray}
where $\hat{\mathbf{t}}_{\mathrm{E}} = [\hat{x}_{\mathrm{E}}, \hat{y}_{\mathrm{E}}]^{\mathrm{T}} \in \mathbb{R}^{2 \times 1} $ denotes the estimated location of the eavesdropper and $\Delta \mathbf{t}_{\mathrm{E}} = [\Delta x_{\mathrm{E}}, \Delta y_{\mathrm{E}}]^{\mathrm{T}} \in \mathbb{R}^{2 \times 1} $ denotes the estimation error of $\hat{\mathbf{t}}_{\mathrm{E}}$.
The estimation error satisfies
\begin{eqnarray} \label{eqn:Delta_condition}
\| \Delta \mathbf{t}_{\mathrm{E}} \|^2 \leq Q_{\mathrm{E}}^2,
\end{eqnarray}
where $Q_{\mathrm{E}}$ is the radius of the uncertain circular region surrounding the estimated location of the eavesdropper.

To facilitate the design of energy-efficient resource allocation,  the system power consumption is defined as follows.
The flight power consumption for rotary-wing UAV at time slot $n$ with respect to (w.r.t.) velocity $\mathbf{v}[n] = [v_x[n],v_y[n]]^{\mathrm{T}} \in \mathbb{R}^{2 \times 1} $ is given by \cite{rotary_wing_power}:
\begin{eqnarray} \label{eqn:p_flight}
P_{\mathrm{flight}}[n] \hspace{-0.02in} &=& \hspace{-0.02in}  P_o \bigg( 1 + \frac{3 \|\mathbf{v} [n]\|^2 }{\Omega^2 r^2} \bigg) + \frac{P_i v_0}{\|  \mathbf{v} [n] \|} \notag\\
&+& \hspace{-0.01in} \frac{1}{2} d_0 \rho s A \|\mathbf{v} [n]\|^3 ,
\end{eqnarray}
where the notations and the physical meanings of the variables in \eqref{eqn:p_flight} are summarized in Table \ref{notations}.
The total power consumption at time slot $n$ in Joules-per-second (J/sec) includes the communication power and flight power consumptions which can be modeled as
\begin{eqnarray} \label{eqn:total_power}
P_{\mathrm{total}}[n]= \sum_{k=1}^{K} \sum_{i=1}^{N_{\mathrm{F}}} \alpha_{k}^i[n] p_{k}^i[n] + P_{\mathrm{C}}+ P_{\mathrm{flight}}[n],
\end{eqnarray}
where $p_{k}^i[n]$ denote the power allocation of user $k$ on subcarrier $i\in\{1,\ldots,N_{\mathrm{F}}\}$ at time slot $n$ and $P_{\mathrm{C}}$ denotes the constant circuit power consumption.
Variable $\alpha_k^i[n]=1$ represents that subcarrier $i$ is assigned to user $k$ at time slot $n$.
Otherwise, $\alpha_k^i[n]=0$.
We assume that both channels from the UAV to users and from the UAV to the eavesdropper are dominated by LoS links.
Thus, the channel power gain between the UAV and user $k$ in time slot $n$ can be characterized by the commonly adopted free-space path loss model \cite{DWK_2018_solar_power_comm,rotary_wing_power,li2018joint} which is given by
\begin{eqnarray}
h_k[n] = \frac{\beta_0}{d_k^2[n]} = \frac{\beta_0} {\|\mathbf{t}_k-\mathbf{t}[n]\|^2 +H^2},
\end{eqnarray}
where $\beta_0$ represents the channel power gain at a reference distance of 1 meter.
The data rate for user $k$ on subcarrier $i$ at time slot $n$ is given by
\begin{eqnarray} \label{eqn:R_k}
R_{k}^i[n]  = W  \alpha_k^i[n]  \log_2 \bigg( 1 + \frac{ p_k^i[n] h_k[n]} {W N_0} \bigg) ,
\end{eqnarray}
where $W$ represents the subcarrier bandwidth and $N_0$ is the power spectral density of the additive white Gaussian noise (AWGN).
On the other hand, the signal-to-noise ratio (SNR) leakage between the UAV and the potential eavesdropper $E$ on subcarrier $i$ for user $k$ at time slot $n$ is given by
\begin{eqnarray} \label{eqn:R_E}
\mathrm{S N R}_{\mathrm{E},k}^i[n] = \frac{p_k^i[n] \beta_0}{W N_0 d_{\mathrm{E}}^2[n] }.
\end{eqnarray}
Thus, the system energy efficiency in bits-per-Joule (bits/J) is defined as
\begin{eqnarray} \label{eqn:ee}
\mathrm{EE}(\mathcal{A},\mathcal{P},\mathcal{T},\mathcal{V}) = \frac{\frac{1}{N} \sum_{k=1}^{K} \sum_{i=1}^{N_{\mathrm{F}}} \sum_{n=1}^{N} R_{k}^i[n]} {\frac{1}{N} \sum_{n=1}^{N} P_{\mathrm{total}}[n]},
\end{eqnarray}
where the user scheduling variable set as $\mathcal{A}=\{\alpha_{k}^i[n], \forall k,i,n\}$, the transmit power variable set as $\mathcal{P}=\{p_{k}^i[n], \forall k,i,n\}$, the UAV's trajectory variable set as $\mathcal{T}=\{\mathbf{t}[n], \forall n\}$, and the velocity variable set as $\mathcal{V}=\{\mathbf{v}[n], \forall n\}$.
\begin{table}[t]\vspace*{2mm}
\caption{Notations and physical meaning for flight power consumption.} \label{notations}
\begin{tabular}{ c | c }
  \hline			
  Notations & Physical meaning \\ \hline
  $\Omega$ & Blade angular velocity in radians/second \\
  $r$ & Rotor radius in meter \\
  $\rho$ & Air density in $\mathrm{kg/m^3}$ \\
  $s$ & Rotor solidity in $\mathrm{m^3}$ \\
  $A$ & Rotor disc area in $\mathrm{m^2}$ \\
  $P_o$ & Blade profile power in hovering status in W\\
  $P_i$ & Induced power in hovering status in W\\
  $v_0$ & Mean rotor induced velocity in forwarding flight in m/s\\
  $d_0$ & Fuselage drag ratio \\
  \hline
\end{tabular}

\end{table}

\section{Problem Formulation}

 The energy-efficient design of user scheduling, power allocation, UAV's trajectory, and flight velocity is formulated as the following optimization problem:
\begin{eqnarray} \label{eqn:ee_optimal}
\underset {\mathcal{A},\mathcal{P},\mathcal{T},\mathcal{V}} {\text{maximize}} && \mathrm{EE}(\mathcal{A},\mathcal{P},\mathcal{T},\mathcal{V}) \\
\mathrm{s.t.}\,\,\mathrm{C1}: &&\hspace{-0.25in} \alpha_{k}^i[n] \in \{0,1\}, \forall k,i,n, \hspace{0.03in}
\mathrm{C2}: \sum_{k=1}^K \alpha_{k}^i[n] \leq 1, \forall i,n, \notag \\
\mathrm{C3}: &&\hspace{-0.25in} p_{k}^i[n] \geq 0, \forall k,i,n, \notag \\
\mathrm{C4}: &&\hspace{-0.25in} \sum_{k=1}^K \sum_{i=1}^{N_{\mathrm{F}}} \alpha_k^i[n] p_k^i[n] \leq P_{\mathrm{peak}}, \forall n, \notag \\
\mathrm{C5} : &&\hspace{-0.25in} P_{\mathrm{total}}[n] \leq  P_{\max}, \forall n , \notag \\
\mathrm{C6}: &&\hspace{-0.25in} \frac{1}{N} \sum_{i=1}^{N_{\mathrm{F}}} \sum_{n=1}^N R_{k}^i[n] \geq R_{\min}, \forall k, \notag \\
\mathrm{C7}: &&\hspace{-0.25in} \underset{ \|\Delta \mathbf{t}_{\mathrm{E}}\| \leq Q_{\mathrm{E}}}{\max} \,\, \mathrm{S N R}_{\mathrm{E},k}^i[n] \leq \Gamma_{\mathrm{th}}, \forall k,i,n, \notag \\
\mathrm{C8}: &&\hspace{-0.25in} \mathbf{t}[0] = \mathbf{t}_0, \hspace{0.81in}
\mathrm{C9}: \mathbf{t}[N] = \mathbf{t}_\mathrm{F}, \notag \\
\mathrm{C10}: &&\hspace{-0.25in} \mathbf{t}[n+1]=\mathbf{t}[n]+\mathbf{v}[n] \tau, n=1,...,N-1, \notag \\
\mathrm{C11}: &&\hspace{-0.25in} \|\mathbf{v}[n]\| \hspace{-0.02in} \leq \hspace{-0.02in} V_{\max}, \hspace{-0.02in} \forall n, \notag \\
\mathrm{C12}: &&\hspace{-0.25in} \|\mathbf{v}[n+1]-\mathbf{v}[n]\| \leq V_{\mathrm{acc}}, n=1,...,N-1. \notag
\end{eqnarray}
Note that $\mathrm{C1}$ and $\mathrm{C2}$ are user scheduling constraints such that each subcarrier at each time slot can be assigned to at most one user\footnote{The extension to non-orthogonal multiple access \cite{Zhiqiang_DC,sun2017optimal} and massive multiple-input multiple-output \cite{zhang2016spectral,Mixed-ADC/DAC} will be considered in our future work.} to avoid multiple access interference.
$\mathrm{C3}$ is the non-negative power constraint.
$P_{\mathrm{peak}}$ in $\mathrm{C4}$ is the peak transmit power at each time slot.
$P_{\max}$ in $\mathrm{C5}$ is the maximum limitation for total power consumption at each time slot.
$R_{\min}$ in $\mathrm{C6}$ denotes the minimum required individual user data rate.
$\Gamma_{\mathrm{th}}$ in $\mathrm{C7}$ is the maximum tolerable SNR threshold for the potential eavesdropper in eavesdropping the information of user $k$ on subcarrier $i$.
Note that constraint $\mathrm{C7}$ takes into account the location uncertainty of the potential eavesdropper.
$\mathrm{C8}$ and $\mathrm{C9}$ indicate the required UAV's initial and final locations, respectively.
$\mathrm{C10}$ draws the connections between the velocity of the UAV and its displacement at two consecutive time slots.
$\mathrm{C11}$ is the UAV's the maximum flight velocity constraint.
$V_{\mathrm{acc}}$ in constraint $\mathrm{C12}$ is the maximum allowable acceleration in a given time slot.
Note that the  flight velocity of a UAV can be expressed as a function of its trajectory for a given constant time slot duration $\tau$.
Yet, expressing the flight power consumption as a function of trajectory would complicate the resource allocation design.
Therefore, we introduce the flight velocity variable $\mathbf{v}[n]$ to simplify the problem formulation.
\begin{Remark}\textup{In the considered system,  secure communication can be guaranteed when $R_{\min}>\log_2(1+\Gamma_{\mathrm{th}})$ holds. In particular, the parameters
$R_{\min}$ and $\Gamma_{\mathrm{th}}$ can be chosen by the system operator to provide flexibility in
designing  resource allocation algorithms for
different applications requiring different levels of communication security and the adopted formulation has been widely adopted, e.g. \cite{JR:Sesure_SWIPT}.}\end{Remark}

\section{Problem Solution}

The formulated problem in \eqref{eqn:ee_optimal} is non-convex, which generally cannot be solved efficiently by conventional convex optimization methods.
To facilitate a low computational complexity design of resource allocation and trajectory, we divide the problem \eqref{eqn:ee_optimal} into two sub-problems and solve them iteratively to achieve a sub-optimal solution using the alternating optimization approach \cite{Alternating}.
In particular, sub-problem 1 aims to optimize the user scheduling $\mathcal{A}$ and the transmit power allocation $\mathcal{P}$ for a given feasible UAV's trajectory $\mathcal{T}$ and flight velocity $\mathcal{V}$.
On the other hand, sub-problem 2 aims to optimize the UAV's trajectory $\mathcal{T}$ and flight velocity $\mathcal{V}$ under a given feasible user scheduling $\mathcal{A}$ and transmit power allocation $\mathcal{P}$.
Now, we first study the solution of sub-problem 1.

\subsection{Sub-problem 1: Optimizing User Scheduling and Transmit Power Allocation}

For a given UAV's trajectory $\mathcal{T} = \{\mathbf{t}[n], \forall n\}$ and flight velocity $\mathcal{V} = \{\mathbf{v}[n], \forall n\}$, we can express sub-problem 1 as the following optimization problem:
\begin{eqnarray} \label{eqn:sub_optimal_1}
\underset {\mathcal{A},\mathcal{P}} {\text{maximize}} && \hspace{-0.2in} \frac{\frac{1}{N} \sum_{k=1}^{K} \sum_{i=1}^{N_{\mathrm{F}}} \sum_{n=1}^{N} R_k^i[n]} { \frac{1}{N} \sum_{n=1}^{N} P_{\mathrm{total}}[n]} \\
\mathrm{s.t.}\,\, && \hspace{-0.2in} \mathrm{C1}-\mathrm{C7}. \notag
\end{eqnarray}
In order to solve sub-problem 1 in \eqref{eqn:sub_optimal_1}, we introduce an auxiliary variable $\tilde{p}_k^i[n] = \alpha_k^i[n] p_k^i[n], \forall k,i,n$,
and the transformed problem is given by
\begin{eqnarray} \label{eqn:sub_optimal_1_1}
\underset {\mathcal{A},\tilde{\mathcal{P}}} {\text{maximize}} && \hspace{-0.2in} \frac{\frac{1}{N} \sum_{k=1}^{K} \sum_{i=1}^{N_{\mathrm{F}}} \sum_{n=1}^{N} \tilde{R}_{k}^i[n]} {\frac{1}{N} \sum_{n=1}^{N} \tilde{P}_{\mathrm{total}}[n]} \\
\hspace{-0.25in} \mathrm{s.t.} \,\, \mathrm{C1} , && \hspace{-0.25in} \mathrm{C2}, \notag \\[-0.5mm]
\widetilde{\mathrm{C3}} \hspace{-0.04in}: && \hspace{-0.25in} \tilde{p}_{k}^i[n] \geq 0, \forall k,i,n, \hspace{0.05in}
\widetilde{\mathrm{C4}} \hspace{-0.04in}: \sum_{k=1}^K \sum_{i=1}^{N_{\mathrm{F}}} \tilde{p}_{k}^i[n] \leq P_{\mathrm{peak}}, \forall n, \notag \\
\widetilde{\mathrm{C5}} \hspace{-0.04in}: && \hspace{-0.25in} \tilde{P}_{\hspace{-0.02in}\mathrm{total}}[n] \hspace{-0.02in} \leq \hspace{-0.03in} P_{\hspace{-0.03in} \max}, \hspace{-0.02in} \forall \hspace{-0.01in}n,
\widetilde{\mathrm{C6}} \hspace{-0.04in}: \hspace{-0.02in} \frac{1}{N} \hspace{-0.03in} \sum_{i=1}^{N_{\mathrm{F}}} \hspace{-0.02in} \sum_{n=1}^N \hspace{-0.03in} \tilde{R}_{k}^i[n] \hspace{-0.03in} \geq \hspace{-0.03in} R_{\min}, \hspace{-0.01in} \forall k, \notag \\
\widetilde{\mathrm{C7}}\hspace{-0.04in}: && \hspace{-0.25in} \tilde{p}_{k}^i[n] \leq \alpha_{k}^i[n] \frac{W N_0 \Gamma_{\mathrm{th}}} {\beta_0} \underset{ \| \Delta \mathbf{t}_{\mathrm{E}}\| \leq Q_{\mathrm{E}} }{\min} \,\, d_{\mathrm{E}}^2[n], \forall k,i,n, \notag
\end{eqnarray}
where $\tilde{\mathcal{P}} = \{\tilde{p}_k^i[n], \forall k,i,n\}$,
\begin{eqnarray}
\hspace{-0.13in} \tilde{R}_k^i[n] \hspace{-0.06in}&=&\hspace{-0.1in} W \alpha_k^i[n] \hspace{-0.01in} \log_2 \hspace{-0.01in} \bigg( \hspace{-0.01in} 1 \hspace{-0.01in} + \hspace{-0.01in} \frac{\tilde{p}_k^i[n] h_k[n]} {W N_0 \alpha_k^i[n]} \hspace{-0.01in} \bigg) \hspace{-0.01in} , \hspace{-0.01in} \forall k,i,n, \, \text{and} \,\, \label{eqn:R_k_tilde} \\
\hspace{-0.13in} \tilde{P}_{\mathrm{total}}[n] \hspace{-0.12in}&=&\hspace{-0.1in} \sum_{k=1}^{K} \sum_{i=1}^{N_{\mathrm{F}}} \tilde{p}_{k}^i[n] + P_{\mathrm{C}}+ P_{\mathrm{flight}}[n], \forall n. \label{eqn:total_power_tilde}
\end{eqnarray}
Note that since the trajectory of the UAV is given for sub-problem 1, the minimum distance between the UAV and the potential eavesdropper is known.
In other words, $\underset{ \| \Delta \mathbf{t}_{\mathrm{E}}\| \leq Q_{\mathrm{E}} }{\min} \,\, {d_{\mathrm{E}}}^2[n]$ is a constant for a given uncertain area of the eavesdropper.
The main obstacle in solving \eqref{eqn:sub_optimal_1_1} arises from the binary user scheduling constraint $\mathrm{C1}$ and the objective function in fractional form.
First, we handle the binary constraint.
In particular, we follow the approach as in \cite{ng_EE}, and relax the subcarrier variable $\alpha_k^i[n]$ such that it is a real value between $0$ and $1$, i.e.,
\begin{eqnarray} \label{eqn:relax_a}
0 \leq \alpha_k^i[n] \leq 1, \forall k,i,n.
\end{eqnarray}
Meanwhile,  the relaxed version of $\alpha_k^i[n]$ serves as a time-sharing factor for  user $k$ in utilizing subcarrier $i$  at time slot $n$.
Note that the relaxation is asymptotically tight even if the number of subcarriers is small, e.g. $8$ subcarriers \cite{ng_EE}.
Then, we tackle the fractional-form objective function.
Let $q_1^*$ be the maximum system energy efficiency of sub-problem 1 which is given by 
\begin{eqnarray} \label{eqn:max_ee}
q_1^* = \frac{R(\mathcal{A}^*,\tilde{\mathcal{P}}^*)} {P(\tilde{\mathcal{P}}^*)} = \underset {\mathcal{A},\tilde{\mathcal{P}}\in \mathcal{F}} {\text{maximize}}\,\, \frac{R(\mathcal{A},\tilde{\mathcal{P}})} {P(\tilde{\mathcal{P}})},
\end{eqnarray}
where $\mathcal{A}^*$ and $\tilde{\mathcal{P}}^*$ are the sets of the optimal user scheduling and power allocation, respectively.
$\mathcal{F}$ is the feasible solution set spanned by constraints $\mathrm{C1}$--$\mathrm{C7}$.
Now, by applying the fractional programming theory \cite{ng_EE},  the objective function of \eqref{eqn:sub_optimal_1_1} from a fractional form can be equivalent transformed into a subtractive form.
More importantly, the optimal value of $q_1^*$ can be achieved if and only if \vspace{-0.3mm}
\begin{eqnarray} \label{eqn:dinc}
\underset {\mathcal{A},\tilde{\mathcal{P}}\in \mathcal{F}} {\text{maximize}} &&  R(\mathcal{A},\tilde{\mathcal{P}}) - q_1^* P(\tilde{\mathcal{P}}) \\[-1.3mm]
= &&  R(\mathcal{A}^*,\tilde{\mathcal{P}}^*) - q_1^* P(\tilde{\mathcal{P}}^*) = 0, \notag
\end{eqnarray}
for $R(\mathcal{A},\tilde{\mathcal{P}}) \geq 0$ and $P(\tilde{\mathcal{P}}) > 0$.
Therefore, we can apply the iterative Dinkelbach method \cite{dinkelbach} to solve \eqref{eqn:sub_optimal_1_1}.
In particular, for the $g^{\mathrm{Algo1}}$-th iteration for sub-problem 1 and a given intermediate value $q_1^{(g^{\mathrm{Algo1}})}$, we need to solve a convex optimization as follows: \vspace{-1mm}
\begin{eqnarray} \label{eqn:sub_optimal_1_2}
\hspace{-0.2in} \{\underline{\mathcal{A}},\underline{\tilde{\mathcal{P}}}\}= \arg \, \underset {\mathcal{A},\tilde{\mathcal{P}}} {\text{maximize}} && \hspace{-0.2in} \frac{1}{N} \sum_{k=1}^{K} \sum_{i=1}^{N_{\mathrm{F}}} \sum_{n=1}^{N} \tilde{R}_{k}^i[n] \notag \\[-0.8mm]
&& \hspace{-0.2in} -  q_1^{(g^{\mathrm{Algo1}})} \frac{1}{N}  \sum_{n=1}^{N} \tilde{P}_{\mathrm{total}}[n] \\[-1mm]
\mathrm{s.t.}\,\, \mathrm{C2} \hspace{0.02in} , && \hspace{-0.22in} \widetilde{\mathrm{C3}} - \widetilde{\mathrm{C7}}, \notag \\
\widetilde{\mathrm{C1}} \hspace{-0.02in}: && \hspace{-0.2in} 0 \leq \alpha_k^i[n] \leq 1, \forall k,i,n, \notag
\end{eqnarray}
where ${\underline{\mathcal{A}},\underline{\tilde{\mathcal{P}}}}$ is the optimal solution of \eqref{eqn:sub_optimal_1_2} for a given $q_1^{(g^{\mathrm{Algo1}})}$.
Then, the intermediate energy efficiency value $q_1^{(g^{\mathrm{Algo1}})}$ should be updated as $q_1^{(g^{\mathrm{Algo1}})}=\frac{R(\underline{\mathcal{A}},\underline{\tilde{\mathcal{P}}})} {P(\underline{\tilde{\mathcal{P}}})}$ for each iteration of the Dinkelbach method until convergence\footnote{Note that the convergence of the Dinkelbach method is guaranteed if the problem in \eqref{eqn:sub_optimal_1_2} can be solved optimally \cite{dinkelbach}.}.
In the following, we discuss the solution development for solving \eqref{eqn:sub_optimal_1_2}.
Since problem \eqref{eqn:sub_optimal_1_2} is jointly convex w.r.t. user scheduling $\mathcal{A}$ as well as transmit power allocation $\tilde{\mathcal{P}}$.
Also, it satisfies the Slater's constraint qualification.
Therefore, the strong duality holds and the duality gap is zero.
Hence, solving the dual problem is equivalent to solving the primal problem of sub-problem 1 in \eqref{eqn:sub_optimal_1_2}.
Although we can directly solve \eqref{eqn:sub_optimal_1_2} via numerical convex program solvers, e.g. CVX, it does not shed light on important system design insights such as the impact of optimization variables on the system performance.
To this end, we focus on the resource allocation design for solving the dual problem.
Now, we first derive the Lagrangian function of  \eqref{eqn:sub_optimal_1_2}: \vspace{-4mm}

\begin{eqnarray} \label{eqn:lagrangian}
\hspace{-0.27in} && \mathcal{L}({\bm\eta},{\bm\varphi},{\bm\theta},{\bm\omega},{\bm\varepsilon},\mathcal{A},\tilde{\mathcal{P}}) \\[-0.4mm]
\hspace{-0.25in} && = \hspace{-0.015in} \sum_{k=1}^K \hspace{-0.015in} \frac{(1\hspace{-0.015in} + \hspace{-0.015in} \omega_{k})}{N} \hspace{-0.02in} \sum_{i=1}^{N_{\mathrm{F}}} \hspace{-0.015in} \sum_{n=1}^N \hspace{-0.015in} \tilde{R}_{k}^i \hspace{-0.005in} [n] \hspace{-0.015in} - \hspace{-0.015in} \big( \hspace{-0.015in} \frac{q_1}{N} \hspace{-0.005in} + \hspace{-0.005in} \theta_n \hspace{-0.02in} \big) \hspace{-0.005in} \tilde{P}_{\mathrm{total}}[n] \hspace{-0.015in} + \hspace{-0.001in} \theta_n P_{\max} \notag \\[-0.4mm]
\hspace{-0.25in} && - \hspace{-0.015in} \sum_{i=1}^{N_{\mathrm{F}}} \hspace{-0.015in} \sum_{n=1}^N \hspace{-0.015in} \eta_{i,n} \bigg( \hspace{-0.03in} \sum_{k=1}^K \hspace{-0.02in} \alpha_{k}^i[n] \hspace{-0.02in} - \hspace{-0.02in} 1 \hspace{-0.03in}  \bigg) \hspace{-0.03in} - \hspace{-0.04in} \sum_{n=1}^N \hspace{-0.02in} \varphi_n \hspace{-0.02in} \bigg( \hspace{-0.015in}  \sum_{k=1}^K \hspace{-0.015in} \sum_{i=1}^{N_{\mathrm{F}}} \hspace{-0.01in} \tilde{p}_{k}^i[n] \hspace{-0.025in} - \hspace{-0.02in}  P_{\mathrm{peak}} \hspace{-0.03in} \bigg) \notag \\[-0.4mm]
\hspace{-0.25in} && - \hspace{-0.015in} \sum_{k=1}^K \hspace{-0.015in} \sum_{i=1}^{N_{\mathrm{F}}} \hspace{-0.015in} \sum_{n=1}^N \hspace{-0.015in} \varepsilon_{k,i,n} \hspace{-0.01in} \bigg( \hspace{-0.01in} \tilde{p}_{k}^i[n] \hspace{-0.02in} - \hspace{-0.02in} \alpha_{k}^i[n] \frac{W N_0 \Gamma_{\mathrm{th}}} {\beta_0} \underset{\| \Delta \mathbf{t}_{\mathrm{E}}\| \leq Q_{\mathrm{E}} }{\min} \,\, d_{\mathrm{E}}^2[n] \hspace{-0.02in} \bigg) \notag \\[-0.4mm]
\hspace{-0.25in} && - \hspace{-0.025in} \sum_{k=1}^K \hspace{-0.01in} \omega_{k} \hspace{-0.01in} R_{\min} , \hspace{-0.02in} \notag
\end{eqnarray}
where ${\bm\eta}=\{\eta_{i,n},\forall i,n\}$, ${\bm\varphi}=\{\varphi_n, \forall n\}$, ${\bm\theta}=\{\theta_n,\forall n\}$, ${\bm\omega}=\{\omega_{k},\forall k\}$, and ${\bm\varepsilon}=\{\varepsilon_{k,i,n},\forall k,i,n\}$ denote the Lagrange multipliers for constraints $\mathrm{C2}$, $\widetilde{\mathrm{C4}}$, $\widetilde{\mathrm{C5}}$, $\widetilde{\mathrm{C6}}$, and $\widetilde{\mathrm{C7}}$, respectively.
Constraints $\mathrm{C1}$ and $\widetilde{\mathrm{C3}}$ will be considered in the Karush-Kuhn-Tucker (KKT) conditions when deriving the optimal solution in the following.
Then, the dual problem of \eqref{eqn:sub_optimal_1_2} is given by
\begin{eqnarray} \label{eqn:dual_problem}
\mathcal{D} = \underset{{\bm\eta},{\bm\varphi},{\bm\theta},{\bm\omega},{\bm\varepsilon} \geq 0}{\text{minimize}} \,\, \underset{\mathcal{A},\tilde{\mathcal{P}}}{\text{maximize}} \,\, \mathcal{L}({\bm\eta},{\bm\varphi},{\bm\theta},{\bm\omega},{\bm\varepsilon},\mathcal{A},\tilde{\mathcal{P}}).
\end{eqnarray}
Subsequently, the dual problem is solved iteratively via dual decomposition. In particular, the dual problem is decomposed into two nested layers:
Layer 1, maximizing the Lagrangian over user scheduling $\mathcal{A}$ and power allocation $\tilde{\mathcal{P}}$ in \eqref{eqn:dual_problem}, given the Lagrange multipliers $\bm\eta$, $\bm\varphi$, $\bm\theta$, $\bm\omega$, and $\bm\varepsilon$;
Layer 2, minimizing the Lagrangian function over $\bm\eta$, $\bm\varphi$, $\bm\theta$, $\bm\omega$, and $\bm\varepsilon$ in \eqref{eqn:dual_problem}, for a given user scheduling $\mathcal{A}$ and power allocation $\tilde{\mathcal{P}}$.
\emph{Solution of Layer 1 (Power Allocation and User Scheduling):}
We assume that ${\alpha_{k}^i}^*[n]$ and $\tilde p{^i_k}^*[n]$ denote the optimal solutions of sub-problem 1. Then,
the optimal power allocation for user $k$ on subcarrier $i$ at time slot $n$ is given by
\begin{eqnarray} \label{eqn:p_sub_1}
\tilde p{^i_k}^{\hspace{-0.02in}*}\hspace{-0.02in}[n] = \alpha_{k}^i[n] {p_{k}^i}^{\hspace{-0.02in}*} \hspace{-0.02in} [n] = \alpha_{k}^i \hspace{-0.01in} [n] \hspace{-0.01in} \bigg[ \hspace{-0.01in} \frac{1 + \omega_{k}}{\Theta_{k,i,n} \ln \hspace{-0.01in} 2} \hspace{-0.02in} - \hspace{-0.02in} \frac{1}{h_{\hspace{-0.01in}k} \hspace{-0.01in} [n]} \hspace{-0.01in} \bigg]^+ \hspace{-0.09in} , \hspace{-0.015in} \forall \hspace{-0.005in} k \hspace{-0.005in},\hspace{-0.01in}i \hspace{-0.005in}, \hspace{-0.01in} n \hspace{-0.005in}, \hspace{-0.04in}
\end{eqnarray}
where $\Theta_{k,i,n} = q_1+N (\varepsilon_{n,k,i}+\theta_n+\varphi_n) $.
The optimal power allocation in \eqref{eqn:p_sub_1} is the classic multiuser water-filling solution.
The water-levels for different users, i.e.,  $\frac{1 + \omega_{k}}{\Theta_{k,i,n} \ln 2}$, are generally different on  different subcarrier $i$ and time slot $n$.
In particular, on one hand, the Lagrange multiplier $\omega_{k}$ forces the UAV to increase the transmit power to satisfy the minimum required individual user data rate $R_{\min}$ of the system.
On the other hand, the Lagrange multiplier  $\varepsilon_{n,k,i}$ adjusts the water-level such that the maximum SNR leakage constraint in $\mathrm{C7}$ can be satisfied.
Besides, to find the optimal subcarrier allocation, we take the derivative of the Lagrangian function w.r.t. $\alpha_{k}^i[n]$ which yields
\begin{eqnarray} \label{eqn:diff_L_a=0}
M_k^i[n] \hspace{-0.13in}&=&\hspace{-0.14in} \frac{(1 \hspace{-0.02in} + \hspace{-0.02in} \omega_{k})}{N} \hspace{-0.02in} \bigg[ \hspace{-0.02in} \log_2 \hspace{-0.01in} ( \hspace{-0.005in} 1 \hspace{-0.02in} + \hspace{-0.02in} p_{k}^i\hspace{-0.01in}[n] h_k\hspace{-0.01in}[n] \hspace{-0.005in} ) \hspace{-0.02in} - \hspace{-0.02in} \frac { p_{k}^i[n] h_k[n]} { ( 1 \hspace{-0.02in} + \hspace{-0.02in} p_{k}^i[n] h_k[n] ) \hspace{-0.01in} \ln 2} \hspace{-0.02in} \bigg] \notag \\[-0.5mm]
\hspace{-0.14in}&-&\hspace{-0.14in} \eta_{i,n} + \varepsilon_{k,i,n} \frac{W N_0 d_{\mathrm{E}}^2[n] \Gamma_{\mathrm{th}}} {\beta_0}.
\end{eqnarray}
In fact, $M_k^i[n] \geq 0$ denotes the marginal benefit of the system performance improvement when subcarrier $i$ is allocated to user $k$ at time slot $n$.
As \eqref{eqn:diff_L_a=0} is independent of $\alpha_k^i[n]$, due to constraint $\mathrm{C2}$, the optimal user scheduling for each subcarrier $i$ and time slot $n$ is given by
\begin{eqnarray} \label{eqn:a_sub_1}
{\alpha_{k}^i}^*[n]= \left\{
  \begin{array}{ll}
  1, & k^*=\underset{k} \max (M_k^i[n]), \\
  0, & \mathrm{otherwise},
  \end{array}
\right.
\forall i,n.
\end{eqnarray}
\emph{Solution of Layer 2 (Master Problem):}
To solve Layer 2 master minimization problem in \eqref{eqn:dual_problem}, the gradient method is adopted and the Lagrange multipliers can be updated by
\begin{eqnarray}
\hspace{-0.2in} \varphi_n (g \hspace{-0.02in} + \hspace{-0.02in} 1) \hspace{-0.11in}&=&\hspace{-0.14in} \bigg[ \hspace{-0.01in}  \varphi_n(g) \hspace{-0.03in} - \hspace{-0.03in} \lambda_1(g) \hspace{-0.02in} \times \hspace{-0.025in} \bigg( \hspace{-0.035in} P_{\mathrm{peak}} \hspace{-0.02in} - \hspace{-0.04in} \sum_{k=1}^K \hspace{-0.015in} \sum_{i=1}^{N_{\mathrm{F}}} \tilde{p}_{k}^i[n] \hspace{-0.02in} \bigg) \hspace{-0.02in} \bigg]^+ \hspace{-0.07in}, \hspace{-0.03in} \forall n \hspace{-0.005in} , \label{eqn:varphi} \\[-0.3mm]
\hspace{-0.2in} \theta_n (g \hspace{-0.02in} + \hspace{-0.02in} 1) \hspace{-0.11in}&=&\hspace{-0.14in} \bigg[ \theta_n(g) - \lambda_2(g) \times \bigg( P_{\max} - \tilde{P}_{\mathrm{total}}[n] \bigg) \bigg]^+ \hspace{-0.08in} , \forall n, \\[-0.3mm]
\hspace{-0.2in} \omega_{k}(g \hspace{-0.02in} + \hspace{-0.02in} 1) \hspace{-0.11in}&=&\hspace{-0.14in} \bigg[ \omega_{k}\hspace{-0.005in} (\hspace{-0.005in} g\hspace{-0.005in} ) \hspace{-0.03in} + \hspace{-0.03in} \lambda_3\hspace{-0.005in}(g) \hspace{-0.02in} \times \hspace{-0.03in} \bigg( \hspace{-0.03in} R_{\hspace{-0.005in}\min} \hspace{-0.04in} - \hspace{-0.03in} \frac{1}{N} \hspace{-0.04in} \sum_{i=1}^{N_{\mathrm{F}}} \hspace{-0.015in} \sum_{n=1}^N \hspace{-0.025in} \tilde{R}_{\hspace{-0.01in}k}^i \hspace{-0.01in} [\hspace{-0.005in} n \hspace{-0.005in}] \hspace{-0.035in} \bigg) \hspace{-0.035in} \bigg]^+ \hspace{-0.07in} , \hspace{-0.03in} \forall \hspace{-0.005in} k\hspace{-0.005in}, \label{eqn:omega}
\end{eqnarray}
\begin{eqnarray}
\hspace{-0.2in} \varepsilon_{\hspace{-0.005in} k \hspace{-0.005in} , \hspace{-0.005in} i \hspace{-0.005in} , \hspace{-0.005in} n} \hspace{-0.01in} (\hspace{-0.01in} g \hspace{-0.01in}  + \hspace{-0.01in}  1 \hspace{-0.01in} ) \hspace{-0.14in}&=&\hspace{-0.14in} \bigg[ \varepsilon_{k,i,n}(g) + \lambda_4(g) \times \bigg( \tilde{p}_{k}^i[n] \hspace{-0.04in} \notag \\[-0.5mm]
\hspace{-0.1in}&-&\hspace{-0.1in} \alpha_{k}^i[n] \frac{W  N_0 \Gamma_{\mathrm{th}}} {\beta_0} \underset{\| \Delta \mathbf{t}_{\mathrm{E}}\| \leq Q_{\mathrm{E}} }{\min} \, d_{\mathrm{E}}^2[n] \bigg)  \hspace{-0.02in} \bigg]^+ \hspace{-0.08in} , \hspace{-0.01in} \forall k,\hspace{-0.01in} i, \hspace{-0.01in}n,  \label{eqn:varepsilon}
\end{eqnarray}
where $g \geq 0$ is the iteration index for sub-problem 1 and $\lambda_u(g)$, $u \in \{1,\ldots,4\}$ are step sizes satisfying the infinite travel condition \cite{Ng_L}.
Then, the updated Lagrangian multipliers in \eqref{eqn:varphi}--\eqref{eqn:varepsilon} are used for solving the Layer 1 sub-problem in \eqref{eqn:dual_problem} via updating the resource allocation policies \cite{Ng_L}.
Since the user scheduling and power allocation variables are finite and non-decreasing over iterations for solving the problem, the convergence of the proposed algorithm to the optimal solution of sub-problem 1 is guaranteed.
The proposed Algorithm for sub-problem 1 is summarized in $\textbf{Algorithm \ref{alg_sub_1}}$.
\begin{table}[t]\vspace*{-3mm}
\begin{algorithm} [H]
\caption{Proposed Algorithm for Solving Sub-problem 1} \label{alg_sub_1}
\begin{algorithmic} [1]
\STATE Initialize the maximum number of iterations $G_{\max}^{\mathrm{Algo1}}$
\STATE Set the energy efficiency $q_1^{(0)}=0$ and the iteration index $g^{\mathrm{Algo1}}=0$

\REPEAT[Main Loop]
  \STATE Solve \eqref{eqn:sub_optimal_1_2} for a given $q_1^{(g^{\mathrm{Algo1}})}$ and obtain resource allocation $\{\underline{\mathcal{A}}^{(g^{\mathrm{Algo1}})}, \underline{\tilde{\mathcal{P}}}^{(g^{\mathrm{Algo1}})}\}$

  \IF{$R(\underline{\mathcal{A}}^{(g^{\mathrm{Algo1}})}, \underline{\tilde{\mathcal{P}}}^{(g^{\mathrm{Algo1}})}) - q_1^{(g^{\mathrm{Algo1}})} P(\underline{\tilde{\mathcal{P}}}^{(g^{\mathrm{Algo1}})}) < \epsilon$}
  \STATE  $\mbox{Convergence}=\,$\TRUE
  \RETURN  $\{{\alpha_k^i}[n], {p_k^i}[n]\} = \{ \underline{\mathcal{A}}^{(g^{\mathrm{Algo1}})}, \underline{\tilde{\mathcal{P}}}^{(g^{\mathrm{Algo1}})} \}$ and
  \STATE $q_1 = \frac{R(\underline{\mathcal{A}}^{(g^{\mathrm{Algo1}})}, \underline{\tilde{\mathcal{P}}}^{(g^{\mathrm{Algo1}})})} {P(\underline{\tilde{\mathcal{P}}}^{(g^{\mathrm{Algo1}})})}$

\ELSE

\STATE Set $g^{\mathrm{Algo1}} = g^{\mathrm{Algo1}} + 1$
\STATE $q_1^{(g^{\mathrm{Algo1}})} = \frac{R(\underline{\mathcal{A}}^{(g^{\mathrm{Algo1}})}, \underline{\tilde{\mathcal{P}}}^{(g^{\mathrm{Algo1}})})} {P(\underline{\tilde{\mathcal{P}}}^{(g^{\mathrm{Algo1}})})}$
\STATE  Convergence $=$ \FALSE
\ENDIF

\UNTIL {Convergence $=$ \TRUE $\,$or $g^{\mathrm{Algo1}} = G_{\max}^{\mathrm{Algo1}}$}
\end{algorithmic}
\end{algorithm}
\vspace*{-8mm}
\end{table}

\subsection{Sub-problem 2: Optimizing UAV's Trajectory and Flight Velocity}

For given user scheduling $\mathcal{A} = \{ {\alpha_k^i}[n], \forall k,i,n \}$ and transmit power allocation $\mathcal{P} = \{ {p_k^i}[n], \forall k,i,n\}$, we can express sub-problem 2 as \vspace*{-2mm}
\begin{eqnarray} \label{eqn:sub_optimal_2_initial}
\underset {\mathcal{T},\mathcal{V}} {\text{maximize}} &&  \frac{\frac{1}{N} \sum_{k=1}^{K} \sum_{i=1}^{N_{\mathrm{F}}} \sum_{n=1}^{N} R_k^i[n]} {\frac{1}{N} \sum_{n=1}^{N} P_{\mathrm{total}}[n]} \\
\mathrm{s.t.}\,\, && \mathrm{C5} - \mathrm{C12}. \notag
\end{eqnarray}
The problem in \eqref{eqn:sub_optimal_2_initial} is non-convex and non-convexity arises from the objective function and constraint $\mathrm{C7}$.
To facilitate the derivation of solution,
we introduce a slack variables $u_k[n]$ to transform the problem into the following equivalent form: \vspace*{-0.5mm}
\begin{eqnarray} \label{eqn:sub_optimal_2}
\underset {\mathcal{T},\mathcal{V}, \mathcal{U}} {\text{maximize}} \hspace{-0.2in}&&  \frac{\frac{1}{N} \sum_{k=1}^K \sum_{i=1}^{N_{\mathrm{F}}} \sum_{n=1}^N \bar{R}_k^i[n] } {\frac{1}{N} \sum_{n=1}^{N} P_{\mathrm{total}}[n]} \\
\mathrm{s.t.}\,\, \mathrm{C5} \hspace{0.03in},\hspace{-0.2in}&& \mathrm{C8} - \mathrm{C12}, \notag \\[-0.5mm]
\overline{\mathrm{C6}}: \hspace{-0.2in}&& \frac{1}{N} \sum_{i=1}^{N_{\mathrm{F}}} \sum_{n=1}^N \bar{R}_k^i[n] \geq R_{\min}, \forall k, \notag \\[-1.5mm]
\overline{\mathrm{C7}}: \hspace{-0.2in}&& \underset{\Delta \mathbf{t}_{\mathrm{E}} }{\text{minimize}} \,\, \| \hat{\mathbf{t}}_{\mathrm{E}} \hspace{-0.02in} + \hspace{-0.02in} \Delta \mathbf{t}_{\mathrm{E}} \hspace{-0.02in} - \hspace{-0.02in} \mathbf{t}[n] \|^2 \hspace{-0.02in} + \hspace{-0.02in} H^2 \hspace{-0.02in} \geq \hspace{-0.02in} \frac{\gamma_k^i[n]} { \Gamma_{\mathrm{th}}} \hspace{-0.02in} , \hspace{-0.02in} \forall i \hspace{-0.01in} , \hspace{-0.01in} k \hspace{-0.01in} , \hspace{-0.01in} n \hspace{-0.01in} , \notag \\[-0.5mm]
\mathrm{C13}: \hspace{-0.2in}&& \|\mathbf{t}_k-\mathbf{t}[n]\|^2 + H^2 \leq u_k[n], \forall k,n, \notag
\end{eqnarray}
where $\mathcal{U}=\{u_{k}[n], \forall k,n\}$, \vspace*{-1.5mm}
\begin{eqnarray}
\bar{R}_k^i[n] &=&  W {\alpha_{k}^i}[n] \log_2 \bigg( 1+ \frac{\gamma_k^i[n]} {u_k[n]} \bigg), \,\,\mathrm{and} \\[-1mm]
\gamma_k^i[n] &=& \frac{{p_k^i}[n] \beta_0} {W N_0}.
\end{eqnarray}
It can be proved that \eqref{eqn:sub_optimal_2_initial} and \eqref{eqn:sub_optimal_2} are equivalent as the inequality constraint $\mathrm{C13}$ is active at optimal solution of \eqref{eqn:sub_optimal_2}.
Then, we handle the location uncertainty of the eavesdropper  by rewriting constraint $\overline{\mathrm{C7}}$ as: \vspace*{-0.5mm}
\begin{eqnarray}
\underset{ \| \Delta \mathbf{t}_{\mathrm{E}}\| \leq Q_{\mathrm{E}} }{\max}  - \| \hat{\mathbf{t}}_{\mathrm{E}} + \Delta \mathbf{t}_{\mathrm{E}} - \mathbf{t}[n]\|^2 - H^2 + \frac{\gamma_k^i[n]} { \Gamma_{\mathrm{th}}} \leq 0. \label{eqn:d_e_t_e_leq_0}
\end{eqnarray}
Note that the location uncertainty introduces an infinite number of constraints in $\overline{\mathrm{C7}}$.
To circumvent the difficulty, we apply the $\mathcal{S}$-Procedure \cite{cui2018robust} and transform $\overline{\mathrm{C7}}$ into a finite number of linear matrix inequalities (LMIs) constraints.
In particular,  if there exists a variable $\psi[n] \geq 0$ such that
\begin{eqnarray} \label{eqn:S}
\Phi (\mathbf{t}[n], \psi[n]) \succeq \mathbf{0}, \forall n,
\end{eqnarray}
holds, where
\begin{eqnarray}
\Phi (\mathbf{t}[n], \psi[n]) = \left[ \begin{array}{ccc}
(\psi[n] +1) \mathbf{I}_2 & \mathbf{t}[n] - \hat{\mathbf{t}}_{\mathrm{E}} \\
( \mathbf{t}[n] - \hat{\mathbf{t}}_{\mathrm{E}})^{\mathrm{T}} & -\psi[n] Q_{\mathrm{E}}^2 + c[n]
\end{array}
\right]
\end{eqnarray}
and
\begin{eqnarray} \label{eqn:c}
c[n] = \|\mathbf{t}[n]\|^2 - 2 \|\hat{\mathbf{t}}_{\mathrm{E}}^{\mathrm{T}} \mathbf{t}[n]\| + \| \hat{\mathbf{t}}_{\mathrm{E}}\|^2 +H^2 - \frac{\gamma_k^i[n]} { \Gamma_{\mathrm{th}}},
\end{eqnarray}
then the implication \eqref{eqn:S}$\Rightarrow$\eqref{eqn:d_e_t_e_leq_0} holds.

Note that $c[n]$ in constraint \eqref{eqn:S} is non-convex.
To design a tractable resource allocation, the successive convex approximation (SCA) is applied \cite{EE_fixed_wing,Zhang2018Securing,wei2018multi}.
In particular, for a given feasible solution $\mathbf{t}^{(j^{\mathrm{Algo2}})}[n]$ in the $j^{\mathrm{Algo2}}$-th main loop iteration for sub-problem 2, since $c[n] \geq \tilde{c}^{(j^{\mathrm{Algo2}})}[n]$, we obtain a lower bound of equation \eqref{eqn:S}, which is given by
\begin{eqnarray} \label{eqn:SS}
\widetilde{\Phi}^{(j^{\mathrm{Algo2}})} (\mathbf{t}[n], \psi[n]) \succeq \mathbf{0}, \forall n,
\end{eqnarray}
where \vspace*{-0.5mm}
\begin{eqnarray}
&& \widetilde{\Phi}^{(j^{\mathrm{Algo2}})} (\mathbf{t}[n], \psi[n])  \notag \\
&& = \left[  \begin{array}{ccc}
(\psi[n] +1) \mathbf{I}_2 & \mathbf{t}[n] - \hat{\mathbf{t}}_{\mathrm{E}} \\
(\mathbf{t}[n] - \hat{\mathbf{t}}_{\mathrm{E}})^{\mathrm{T}} & - \psi[n] Q_{\mathrm{E}}^2 + \tilde{c}^{(j^{\mathrm{Algo2}})}[n]
\end{array}
\right]  ,
\end{eqnarray}
and \vspace*{-0.3mm}
\begin{eqnarray}
\tilde{c}^{(j^{\mathrm{Algo2}})} [n] &= & \|\hat{\mathbf{t}}_{\mathrm{E}} \|^2 +  2 \mathbf{t}^{\mathrm{T}} [n] \mathbf{t}^{(j^{\mathrm{Algo2}})} [n] - \mathbf{t}^{(j^{\mathrm{Algo2}})} [n] \notag \\[-0.5mm]
&-& 2 \hat{\mathbf{t}}_{\mathrm{E}}^{\mathrm{T}} \mathbf{t} [n]  +  H^2 - \frac{\gamma_{k}^i[n]} { \Gamma_{\mathrm{th}}} .
\end{eqnarray}

\begin{table}[t] \vspace*{-2mm}
\begin{algorithm} [H]
\caption{Proposed Algorithm for Solving Sub-problem 2} \label{alg_sub_2}
\begin{algorithmic} [1]

\STATE Initialize the maximum number of iterations $J_{\max}^{\mathrm{Algo2}}$, $J_{\mathrm{inner},\max}^{\mathrm{Algo2}}$

\STATE Set the energy efficiency $q_2^{(0)}=0$ and the iteration index $j^{\mathrm{Algo2}}=0$

\REPEAT[Main Loop]
\STATE Set the inner loop iteration index $j_{\mathrm{inner}}^{\mathrm{Algo2}} = 0$

  \REPEAT[Inner Loop]
  \STATE Solve the problem in \eqref{eqn:sub_optimal_2_3} for a given $q_2^{j_{\mathrm{inner}}^{\mathrm{Algo2}}}$ and obtain trajectory and velocity $\{\underline{\mathcal{T}}^{(j_{\mathrm{inner}}^{\mathrm{Algo2}})}$, $\underline{\mathcal{U}}^{(j_{\mathrm{inner}}^{\mathrm{Algo2}})}$, $\underline{\mathcal{V}}^{(j_{\mathrm{inner}}^{\mathrm{Algo2}})}$, $\underline{\Upsilon}^{(j_{\mathrm{inner}}^{\mathrm{Algo2}})}\}$
  \IF{$\bar{R}(\underline{\mathcal{U}}^{(j_{\mathrm{inner}}^{\mathrm{Algo2}})}) - q_2^{(j_{\mathrm{inner}}^{\mathrm{Algo2}})} P^{\mathrm{UB}}(\underline{\mathcal{V}}^{(j_{\mathrm{inner}}^{\mathrm{Algo2}})}, \underline{\Upsilon}^{(j_{\mathrm{inner}}^{\mathrm{Algo2}})}) < \epsilon$}
  \STATE  $\mbox{Inner Loop Convergence}=\,$\TRUE
  \RETURN $\{\underline{\mathcal{T}}^{(j_{\mathrm{inner}}^{\mathrm{Algo2}})}$, $\underline{\mathcal{U}}^{(j_{\mathrm{inner}}^{\mathrm{Algo2}})}$,  $\underline{\mathcal{V}}^{(j_{\mathrm{inner}}^{\mathrm{Algo2}})}\}$ and $q_2^{(j_{\mathrm{inner}}^{\mathrm{Algo2}})}$

  \ELSE
  \STATE Set $j_{\mathrm{inner}}^{\mathrm{Algo2}} = j_{\mathrm{inner}}^{\mathrm{Algo2}} +1$ and
  \STATE $q_2^{(j_{\mathrm{inner}}^{\mathrm{Algo2}})} = \frac{\bar{R}(\underline{\mathcal{U}}^{(j_{\mathrm{inner}}^{\mathrm{Algo2}})})} {P^{\mathrm{UB}}(\underline{\mathcal{V}}^{(j_{\mathrm{inner}}^{\mathrm{Algo2}})}, \underline{\Upsilon}^{(j_{\mathrm{inner}}^{\mathrm{Algo2}})})}$
  \STATE  Inner Loop Convergence $=$ \FALSE

  \ENDIF
  \UNTIL{Inner Loop Convergence $=$ \TRUE $\,$or $j_{\mathrm{inner}}^{\mathrm{Algo2}} = J_{\mathrm{inner},\max}^{\mathrm{Algo2}}$}

\IF{$q_2^{(j^{\mathrm{Algo2}})} - q_2^{(j^{\mathrm{Algo2}}-1)} < \epsilon$}
\STATE  $\mbox{Main Loop Convergence}=\,$\TRUE
\RETURN $\{ \mathbf{t}[n], \mathbf{v}[n] \}$ = $\{\underline{\mathcal{T}}^{(j^{\mathrm{Algo2}})}$, $\underline{\mathcal{U}}^{(j^{\mathrm{Algo2}})}$, $\underline{\mathcal{V}}^{(j^{\mathrm{Algo2}})}\}$  and $q_2 = q_2^{(j^{\mathrm{Algo2}})}$
\ELSE
\STATE Set  $j^{\mathrm{Algo2}} = j^{\mathrm{Algo2}} +1$, $\{\underline{\mathcal{T}}^{(j^{\mathrm{Algo2}})}$, $\underline{\mathcal{U}}^{(j^{\mathrm{Algo2}})}$, $\underline{\mathcal{V}}^{(j^{\mathrm{Algo2}})}\} =$ $\{\underline{\mathcal{T}}^{(j_{\mathrm{inner}}^{\mathrm{Algo2}})}$, $\underline{\mathcal{U}}^{(j_{\mathrm{inner}}^{\mathrm{Algo2}})}$,  $\underline{\mathcal{V}}^{(j_{\mathrm{inner}}^{\mathrm{Algo2}})}\}$ and $q_2^{(j^{\mathrm{Algo2}})} = q_2^{(j_{\mathrm{inner}}^{\mathrm{Algo2}})}$
\STATE  Main Loop Convergence $=$ \FALSE
\ENDIF

\UNTIL{Main Loop Convergence $=$ \TRUE $\,$or $j^{\mathrm{Algo2}}=J_{\max}^{\mathrm{Algo2}}$}
\end{algorithmic}
\end{algorithm}
\vspace*{-8mm}
\end{table}

By replacing constraint $\overline{\mathrm{C7}}$ in \eqref{eqn:sub_optimal_2} with \eqref{eqn:SS} results in a smaller feasible solution set and leads to a performance lower bound of the problem in \eqref{eqn:sub_optimal_2} by solving the resulting optimization problem:
\begin{eqnarray} \label{eqn:sub_optimal_2_2}
\underset {\mathcal{T},\mathcal{V},\mathcal{U},{\bm\Psi}} {\text{maximize}} && \hspace{-0.2in} \frac{\frac{1}{N} \sum_{k=1}^K \sum_{i=1}^{N_{\mathrm{F}}} \sum_{n=1}^N \bar{R}_k^i[n] } { \frac{1}{N} \sum_{n=1}^N P_{\mathrm{total}}[n] } \\
\mathrm{s.t.}\,\,\mathrm{C5} \hspace{0.03in} , && \hspace{-0.2in} \overline{\mathrm{C6}}, \mathrm{C8} - \mathrm{C13},  \notag \\
\overline{\overline{\mathrm{C7}}}: && \hspace{-0.2in}  \widetilde{\Phi}^{(j^{\mathrm{Algo2}})} (\mathbf{t}[n], \psi[n]) \succeq \mathbf{0}, \forall n, \,\,
\mathrm{C14}: \psi[n] \geq 0, \forall n, \notag
\end{eqnarray}
where $\bm{\Psi}=\{\psi[n], \forall n\}$.
Next, we handle the objective function.
In particular, both denominator and the numerator are non-convex functions.
Hence, we aim to develop a lower bound of the objective function.
First, we consider the nominator of the objective function. Based on the SCA, we obtain the lower bound of the data rate as \vspace*{-0.5mm}
\begin{eqnarray} \label{eqn:R_lb}
\bar{R}_k^i[n] \hspace{-0.1in} &\geq& \hspace{-0.1in} \bar{R} {_{k,\mathrm{lb}}^i}^{(j^{\mathrm{Algo2}})} [n] = W  {\alpha_k^i}[n]  \log_2 \bigg( 1 + \frac{\gamma_k^i[n]}{u_{k}^{(j^{\mathrm{Algo2}})} [n]} \bigg)  \notag \\[-1mm]
\hspace{-0.1in} &-& \hspace{-0.1in} \frac{W {\alpha_{k}^i} [n] \gamma_{k}^i [n] ( u_{k} [n]  - u_k^{(j^{\mathrm{Algo2}})} [n] )}{u_k^{(j^{\mathrm{Algo2}})} [n] (u_k^{(j^{\mathrm{Algo2}})} [n] + \gamma_k^i[n]) \ln 2} , \forall k, i,n,
\end{eqnarray}
where $u_k^{(j^{\mathrm{Algo2}})}[n]$ denotes the feasible solution for $u_k[n]$ in the $j^{\mathrm{Algo2}}$-th main loop iteration.
Then, we handle  the non-convex power consumption, i.e., the denominator of the objective function, by rewriting it in its equivalent form:
\begin{eqnarray}\label{eqn:equivalent_model}
{P}_{\mathrm{total}}^{\mathrm{Eq}}[n]= \sum_{k=1}^{K} \sum_{i=1}^{N_{\mathrm{F}}} \alpha_{k}^i[n] p_{k}^i[n] + P_{\mathrm{C}}+ \tilde{P}_{\mathrm{flight}}[n],
\end{eqnarray}
where
\begin{eqnarray} \label{eqn:p_ub}
 \tilde{P}_{\mathrm{flight}}[n]=  P_o \bigg( 1 + \frac{3 \|\mathbf{v}[n]\| ^2}{\Omega^2 r^2} \bigg) + \frac{P_i v_0}{\upsilon[n]} +  \frac{1}{2}d_0\rho sA\|\mathbf{v}[n]\|^3  \notag \\[-4mm] \notag
\end{eqnarray}
is a convex function and variable $\upsilon[n]$ is a new slack optimization variable. In particular, $\upsilon[n]$ satisfies the following two constraints:
\begin{eqnarray}
\mathrm{C15}: &&\hspace{-0.24in}  \|\mathbf{v}[n]\| ^2 \geq \upsilon^2[n], \forall n,  \\
\mathrm{C16}: &&\hspace{-0.24in} \upsilon[n] \geq 0, \forall n. 
\end{eqnarray}
Note that the non-convex constraint $\mathrm{C15}$ is active at the optimal solution and hence the power consumption models in \eqref{eqn:total_power} and \eqref{eqn:equivalent_model} are equivalent. Then, by replacing the power consumption model in \eqref{eqn:total_power} with its equivalent form, the nonconvexity of the denominator of the objective function is captured by constraint $\mathrm{C15}$ which is easier to handle.

Since $\|\mathbf{v}[n]\| ^2$ in $\mathrm{C15}$ is convex and differentiable w.r.t. $\mathbf{v}[n]$, we apply the SCA to obtain its lower bound and improve the bound via an iterative algorithm.  Specifically, for any feasible solution in the $j^{\mathrm{Algo2}}$-th main loop iteration $\mathbf{v}^{(j^{\mathrm{Algo2}})}[n]$, we have
\begin{eqnarray} \label{eqn:v_sca}
\|\mathbf{v}[n]\| ^2 \hspace{-0.02in} &\geq& \hspace{-0.02in} \|\mathbf{v}^{(j^{\mathrm{Algo2}})}[n]\| ^2 \hspace{-0.02in} \notag \\
&+& \hspace{-0.02in} 2 [\mathbf{v}^{(j^{\mathrm{Algo2}})}[n]]^{\mathrm{T}}  (\mathbf{v}[n] - \mathbf{v}^{(j^{\mathrm{Algo2}})}[n]). \\[-5mm] \notag
\end{eqnarray}

\begin{table}[t] \vspace*{-2mm}
\begin{algorithm} [H]
\caption{Overall Algorithm for Solving Problem \eqref{eqn:ee_optimal}} \label{alg_total}
\begin{algorithmic} [1]
\STATE Initialize the maximum number of iterations $L_{\max}^{\mathrm{Algo3}}$ and the maximum tolerance $\epsilon \rightarrow 0$
\STATE Set the iteration index $l^{\mathrm{Algo3}}=0$ and the initial trajectory $\{\mathbf{t}[n], \mathbf{v}[n]\}$

\REPEAT
\STATE Using $\textbf{Algorithm \ref{alg_sub_1}}$ obtain the optimal result $q_1$, $\{\alpha_k^i[n], p_k^i[n]\}$
\STATE Using $\textbf{Algorithm \ref{alg_sub_2}}$ obtain the sub-optimal result $q_2$,  $\{\mathbf{t}[n], \mathbf{v}[n]\}$
\IF{$q_2^{(l^{\mathrm{Algo3}})}-q_2^{(l^{\mathrm{Algo3}}-1)} < \epsilon$}
\STATE $\mbox{Convergence}=\,$\TRUE
\RETURN $ {\alpha_k^i}^*[n]=\alpha_k^i[n], {p_k^i}^*[n]=p_k^i[n], \mathbf{t}^*[n]=\mathbf{t}[n], \mathbf{v}^*[n]=\mathbf{v}[n]$, and $q^*=q_2^{(l^{\mathrm{Algo3}})}$
\ELSE
\STATE Set $l^{\mathrm{Algo3}} = l^{\mathrm{Algo3}}+1$
\STATE  Convergence $=$ \FALSE
\ENDIF
\UNTIL{Convergence $=$ \TRUE $\,$or $l^{\mathrm{Algo3}}=L_{\max}^{\mathrm{Algo3}}$}
\end{algorithmic}
\end{algorithm}
\vspace*{-8mm}
\end{table}
Now, we obtain a lower bound of the objective function via replacing the denominator and the numerator of the original objective function in \eqref{eqn:sub_optimal_2_2} by its equivalent form in \eqref{eqn:equivalent_model} and the lower bound of average total data rate in \eqref{eqn:R_lb}, respectively.
Therefore, we can obtain a lower bound performance of the problem in \eqref{eqn:sub_optimal_2_2} via solving the following optimization problem: 
\begin{eqnarray} \label{eqn:sub_optimal_2_changeP}
\underset {\mathcal{T},\mathcal{V},\mathcal{U},{\bm \Psi},{\bm\Upsilon}} {\text{maximize}} &&\hspace{-0.24in} \frac{\frac{1}{N} \sum_{k=1}^{K} \sum_{i=1}^{N_{\mathrm{F}}} \sum_{n=1}^{N} \bar{R}_{k,\mathrm{lb}}^i[n]} {\frac{1}{N} \sum_{n=1}^{N} {P}_{\mathrm{total}}^{\mathrm{Eq}}[n]} \\
\mathrm{s.t.}\,\, \overline{\overline{\mathrm{C7}}} \hspace{0.03in}, &&\hspace{-0.24in} \mathrm{C8} - \mathrm{C14}, \mathrm{C16}, \notag \\
\overline{\mathrm{C5}}: &&\hspace{-0.25in} P_{\mathrm{total}}^{\mathrm{Eq}} [n] \leq P_{\max}, \forall n, \notag \\
\overline{\overline{\mathrm{C6}}} : &&\hspace{-0.25in} \frac{1}{N} \sum_{i=1}^{N_{\mathrm{F}}} \sum_{n=1}^{N}  \bar{R}_{ k,\mathrm{lb}}^i [n]\geq  R_{\min}, \forall k, \notag \\
\overline{\mathrm{C15}}: &&\hspace{-0.24in}  \|\mathbf{v}^{(j^{\mathrm{Algo2}})} \hspace{-0.005in} [n]\| ^2 \hspace{-0.02in} + \hspace{-0.01in} 2 [\mathbf{v}^{(j^{\mathrm{Algo2}})} \hspace{-0.005in} [n]]^{\mathrm{T}} \hspace{-0.01in} (\mathbf{v}[n] \hspace{-0.01in} - \hspace{-0.01in} \mathbf{v}^{(j^{\mathrm{Algo2}})} \hspace{-0.005in} [n]) \hspace{-0.02in} \notag \\
&&\hspace{-0.24in} \geq \hspace{-0.02in} \upsilon^2[n], \hspace{-0.01in} \forall n, \notag \\[-5mm] \notag
\end{eqnarray}
where ${\bm\Upsilon} = \{\upsilon[n], \forall n\}$.
Now, similar to solving sub-problem 1, the optimal value $q_2^*$ of \eqref{eqn:sub_optimal_2_changeP} can be achieved if and only if
\begin{eqnarray}
\underset {\mathcal{T},\mathcal{U},\mathcal{V},{\bm\Psi},{\bm\Upsilon} \in \bar{\mathcal{F}}} {\text{maximize}} &&  \bar{R}(\mathcal{U}) - q_2^* P^{\mathrm{Eq}}(\mathcal{V},\bm\Upsilon) \\
= &&  \bar{R}(\mathcal{U}^*) - q_2^* P^{\mathrm{Eq}}(\mathcal{V}^*,\bm\Upsilon^*) = 0, \notag 
\end{eqnarray}
for $\bar{R}(\mathcal{T},\mathcal{U}) \geq 0$ and $P^{\mathrm{Eq}}(\mathcal{V},\bm\Upsilon) \geq 0$, where $\bar{\mathcal{F}}$ is the feasible solution set for \eqref{eqn:sub_optimal_2_changeP} and $\mathcal{U}^*,\mathcal{V}^*,\bm\Upsilon^*$ are the optimal trajectory, velocity, and new slack variable sets, respectively.
Then, we can apply the iterative Dinkelbach method to solve \eqref{eqn:sub_optimal_2_changeP} and the details of the proposed algorithm is summarized in $\textbf{Algorithm \ref{alg_sub_2}}$. Specifically, in each inner loop iteration, in line 6 of $\textbf{Algorithm \ref{alg_sub_2}}$, we need to solve the following convex optimization problem\footnote{The problem in \eqref{eqn:sub_optimal_2_3} can be easily solved by dual decomposition or numerical convex program solvers. } for a given $\{\mathbf{t}^{(j^{\mathrm{Algo2}})}[n], \mathbf{v}^{(j^{\mathrm{Algo2}})}[n]\}$ and $q_2^{(j^{\mathrm{Algo2}})}$ \vspace*{-0.5mm}
\begin{eqnarray} \label{eqn:sub_optimal_2_3}
\hspace{-0.22in} \{\underline{\mathcal{T}},\underline{\mathcal{U}},\underline{\mathcal{V}},\underline{\Upsilon}\} \hspace{-0.01in} = \hspace{-0.01in} \arg \, \underset {\mathcal{T},\mathcal{V},\mathcal{U},{\bm\Psi},{\bm\Upsilon}} {\text{maximize}} && \hspace{-0.24in} \frac{1}{N} \sum_{k=1}^{K} \sum_{i=1}^{N_{\mathrm{F}}} \sum_{n=1}^{N} \bar{R}_{k,\mathrm{lb}}^i[n] \notag \\
&& \hspace{-0.24in} - q_2^{(j_{\mathrm{inner}}^{\mathrm{Algo2}})} \hspace{-0.005in} \frac{1}{N} \hspace{-0.03in} \sum_{n=1}^{N} \hspace{-0.005in} P_{\mathrm{total}}^{\mathrm{Eq}}[n] \\
\mathrm{s.t.}\,\, && \hspace{-0.24in} \overline{\mathrm{C5}}, \overline{\overline{\mathrm{C6}}}, \overline{\overline{\mathrm{C7}}}, \mathrm{C8} - \mathrm{C16}, \notag
\end{eqnarray}
where $\{\underline{\mathcal{T}},\underline{\mathcal{U}},\underline{\mathcal{V}},\underline{\Upsilon}\}$ is the optimal solution of \eqref{eqn:sub_optimal_2_3} for a given $q_2^{(j_{\mathrm{inner}}^{\mathrm{Algo2}})}$. After the inner loop converges, we further tighten the bounds obtained by  the SCA via updating
$\{\underline{\mathcal{T}}^{(j^{\mathrm{Algo2}})}$, $\underline{\mathcal{U}}^{(j^{\mathrm{Algo2}})}$, $\underline{\mathcal{V}}^{(j^{\mathrm{Algo2}})}\}$  in the main loop, i.e., line 20 of  $\textbf{Algorithm \ref{alg_sub_2}}$. We note that the convergence of the SCA is guaranteed, cf. \cite{EE_fixed_wing}.

\vspace*{-1mm}
\subsection{Overall Algorithm}

The overall proposed iterative algorithms for solving the two sub-problems \eqref{eqn:sub_optimal_1} and \eqref{eqn:sub_optimal_2_initial} are summarized in $\textbf{Algorithm \ref{alg_total}}$.
Since the feasible solution set of \eqref{eqn:ee_optimal} is compact and its objective value is non-decreasing over iterations via solving the sub-problem in \eqref{eqn:sub_optimal_1} and \eqref{eqn:sub_optimal_2_initial} iteratively, the solution of the proposed algorithm is guaranteed to converge to a suboptimal solution \cite{Alternating}.

\vspace*{-1mm}
\section{Numerical Results}

\begin{figure}[t]
  \centering \vspace*{-2mm}
  \includegraphics[width=3.5 in]{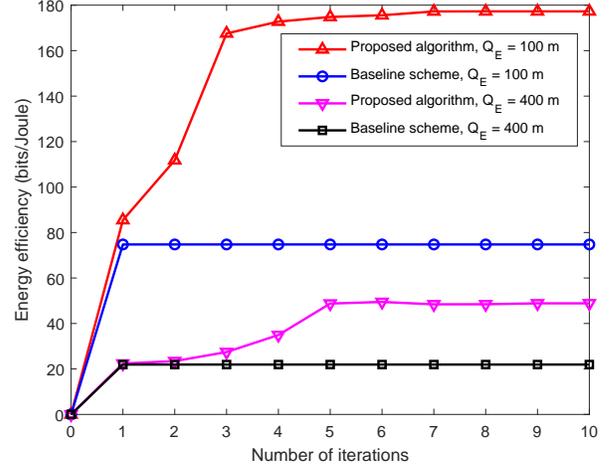} \vspace*{-5.5mm}
  \caption{Energy efficiency versus the number of iterations.}
  \label{EE_Vs_iteration} \vspace*{-3mm}
\end{figure}

In this section, we evaluate the performance of the proposed trajectory and resource allocation design algorithm via simulation.
The simulation setups are summarized in Table \ref{simulation_setting}.
\begin{table}[t]
\caption{Simulation value setting. \cite{EE_fixed_wing}, \cite{rotary_wing_power}} \label{simulation_setting}
\hspace{0.1in} \begin{tabular}{ c | c | c | c }
  \hline			
  Notations & Simulation value & Notations & Simulation value  \\ \hline
  $\Omega$ & 400 radians/second & $ \mathbf{t}_1 $ & $[700;900]$ m \\
  $r$ & 0.5 meter & $ \mathbf{t}_2 $ & $[900;900]$ m \\
  $\rho$ & 1.225 $\mathrm{kg/m^3}$ & $ \mathbf{t}_3 $ & $[900;700]$ m \\
  $s$ & 0.05 & $ \hat{\mathbf{t}}_{\mathrm{E}} $ & $[400;400]$ m \\
  $A$ & 0.79 $\mathrm{m^2}$ & $\mathbf{t}_0$ & $[0;0]$ m \\
  $P_o$ & 580.65 W & $\mathbf{t}_{\mathrm{F}}$ & $[1000;1000]$ m \\
  $P_i$ & 790.67 W& $B$ & 1 MHz \\
  $v_0$ & 7.2 m/s & $W$ & 7.8 kHz \\
  $d_0$ & 0.3 & $N_0$ & -110 dBm/Hz \\
  $K$ & 3 & $P_{\mathrm{C}}$ & $30$ dBm \\
  $N_{\mathrm{F}}$ & 128 & $P_{\max}$ & 65 dBm \\
  $N$ & 50 & $R_{\min}$ & 10 kbits/s \\
  $V_{\max}$ & $50$ m/s & $\Gamma_{\mathrm{th}}$ & -40 dB \\
  $V_{\mathrm{acc}}$ & $5$ m/s & $H$ & 100 m \\
  $\tau$ & 2 second & $G_{\max}^{\mathrm{Algo1}}$ & 10 \\
  $J_{\max}^{\mathrm{Algo2}}$ & 10 & $L_{\max}^{\mathrm{Algo3}}$ & 8 \\
  \hline
\end{tabular}
\vspace*{-2mm}
\end{table}
Fig. \ref{EE_Vs_iteration} illustrates the convergence behavior of the alternating optimization $\textbf{Algorithm \ref{alg_total}}$ for the maximization of the system energy efficiency.
We compare the system performance for different sizes of uncertain areas of the eavesdropper.
For comparison, we also include the performance of a baseline scheme with a straight trajectory between $\mathbf{t}_0=[0;0]$ m and $\mathbf{t}_{\mathrm{F}}=[1000;1000]$ m and a constant cruising velocity.
The peak transmit power is set as $P_{\mathrm{peak}}=1$ W.
It can be seen from Fig. \ref{EE_Vs_iteration} that the system energy efficiency of the proposed algorithm converges to a sub-optimal solution within $8$ iterations.
Thus, in the following results, we set the maximum number of iterations as $8$ to show the performance of the proposed algorithm.
In general, the energy efficiency achieved by the proposed algorithm is superior than that of the baseline scheme.
In fact, the UAV of the proposed algorithm can adjust its transmit power to reduce the chance of information leakage.
Also, it can  avoid the regions and/or reduces the time duration in being close to the eavesdropper by adapting it trajectory. In contrast, in order to guarantee the communication security, the UAV in the baseline scheme would keep its transmit power sufficiently low when it flies close to the eavesdropper.
Moreover, it can be observed that the energy efficiency for a smaller uncertain area of the eavesdropper (e.g. $Q_{\mathrm{E}}=100$ m) is higher than that of the larger uncertain area (e.g. $Q_{\mathrm{E}}=400$ m).
In fact, a larger uncertain area of the eavesdropper would lead to a more stringent security constraint which reduces the flexibility for resource allocation design.

\begin{figure}[t]
  \centering \vspace*{-2mm}
  \includegraphics[width=3.5 in]{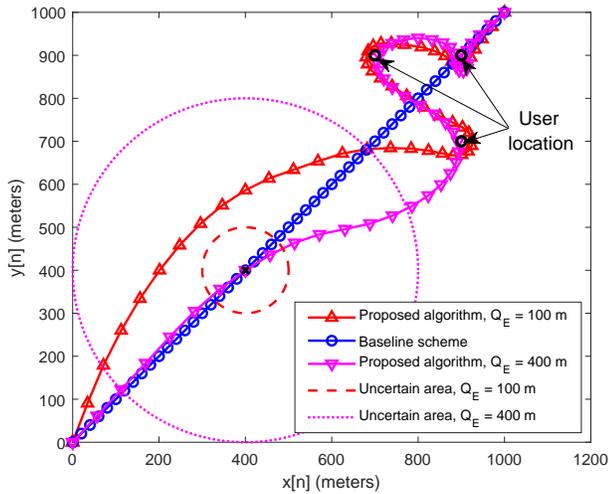}
  \caption{UAV trajectory  with different algorithms and sizes of uncertain areas.} \vspace*{-2mm}
  \label{trajectory}\vspace*{-2mm}
\end{figure}

Fig. \ref{trajectory} shows the UAV's trajectory with the proposed algorithm and the baseline scheme.
The peak transmit power is set as $P_{\mathrm{peak}}=1$ W.
The locations of users and the estimated location of eavesdropper are marked with $\bigcirc$ and $\times$, respectively.
Due to the limited flexibility in optimizing the trajectory, the UAV of the baseline scheme flies directly over the uncertain region, despite the existence of the potential eavesdropper.  Additionally, it can be observed that the proposed algorithm compromises between the energy efficiency and security.
In particular, the UAV of the proposed algorithm would keep a high velocity when it is far away from the users and low velocity when the UAV is close to any desired user.
This behaviour aims to save more time slots for latter when the UAV is close to the users so as to provide higher data rate to the system.
Also, when the uncertain radius of the eavesdropper is small, e.g. $Q_{\mathrm{E}}=100$ m, the UAV of the proposed algorithm tries to keep a distance from the uncertain region while maintains a sufficient transmit power for maximizing the system energy efficiency.
In contrast, when the radius of the potential eavesdropper's uncertain area is sufficiently large, e.g. $Q_{\mathrm{E}}=400$ m, detouring or keeping distance from the uncertain region is not feasible for a given limited time duration.
Thus, the UAV quickly flies through the uncertain region of the eavesdropper to minimize the time duration spending in the region.
Meanwhile, inside the uncertain region,  it only transmits a sufficiently low power to reduce the chance of exceedingly large of signal leakage to the eavesdropper for guaranteeing communication security.
In fact, the UAV allocates a higher amount of energy in cruising than information transmission for leaving the uncertain region as soon as possible.
After the UAV is sufficiently far away from the uncertain region, the transmit power of the UAV would increase again to maximize the system efficiency.

\begin{figure}[t]
  \centering \vspace*{-2mm}
  \includegraphics[width=3.5 in]{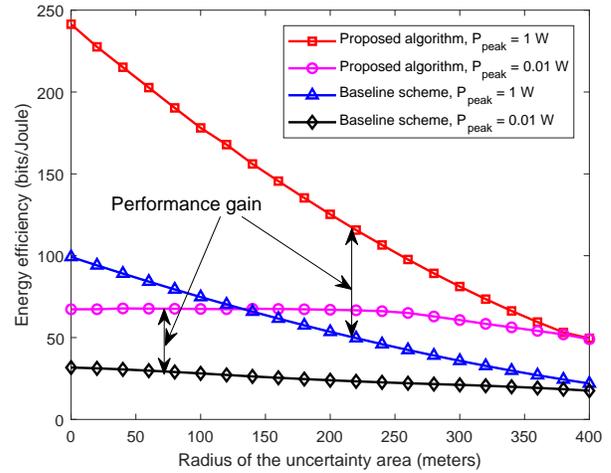}
  \caption{System energy efficiency (bits/Joule) versus the radius of the eavesdropper's uncertain area.} \vspace*{-2mm}
  \label{EE_Vs_radius}\vspace*{-2mm}
\end{figure}

Fig. \ref{EE_Vs_radius} shows the energy efficiency of the considered system versus the radius of the potential eavesdropper's uncertain area.
Although both schemes can guarantee communication security in the considered cases, it can be observed that the energy efficiency of both the proposed algorithm and baseline scheme decreases with the radius of uncertain areas.
Indeed, a larger eavesdropper's uncertain area imposes a more stringent security constraint on the system design, which reduces the flexibility in resource allocation leading to a lower system energy efficiency.
Also, for the proposed algorithm with peak transmit power $P_{\mathrm{peak}}=0.01$ W, the system energy efficiency remains a constant when the radius of the uncertain area is less than $200$ m.
In other words, for a small uncertain area, the system performance is always limited by the small peak power $P_{\mathrm{peak}}$ where the security issue can be handled by trajectory and velocity design.
On the other hand, it can be observed that a large performance gain can be achieved by the proposed algorithm compared to the baseline scheme for a large peak transmit power.
As a matter of fact,  a large peak transmit power offers a higher flexibility for the proposed scheme in allocating the transmit power to achieve a higher system energy efficiency. However, when the peak transmit power is small,  both the trajectory and resource allocation design would become more conservative which reduces the potential performance gain brought by the proposed scheme.
%

\section{Conclusion}

In this paper, we formulated a non-convex energy-efficient maximization problem for secure UAV-OFDMA communication systems via optimizing the resource allocation strategy and the trajectory design.
We proposed a suboptimal algorithm to achieve an efficient solution.
The proposed design enables adaptive velocity and flexible trajectory for UAV which can avoid the potential eavesdropper proactively to guarantee secure communications.
Numerical results demonstrated the fast convergence of the proposed algorithm and the superior performance compared to the  baseline scheme in terms of energy efficiency.
%


\end{document}